\documentclass[aps,pra,reprint,superscriptaddress,amsmath,amssymb,longbibliography,floatfix]{revtex4-2}

\usepackage{graphicx}
\usepackage{bm}
\usepackage{multirow}
\usepackage{xcolor}
\usepackage{url}
\usepackage[hidelinks]{hyperref}

\graphicspath{{./pics/}}

\newcommand{\ket}[1]{|#1\rangle}

\newcommand{\Var}{\operatorname{Var}}
\newcommand{\Bias}{\operatorname{Bias}}
\newcommand{\MSE}{\operatorname{MSE}}
\newcommand{\Fcal}{\mathcal{F}}
\newcommand{\Rcal}{\mathcal{R}}

\newcommand{\Np}{N_{\rm sh}}

\begin{document}

\title{Hybrid physical/logical zero-noise extrapolation with limited logical executions}

\author{Danila Babukhin}
\email{dv.babukhin@gmail.com}
\affiliation{Dukhov Research Institute of Automatics (VNIIA), Moscow 127055, Russia}

\author{Walter Pogosov}
\affiliation{Dukhov Research Institute of Automatics (VNIIA), Moscow 127055, Russia}
\affiliation{Moscow Institute of Physics and Technology, Dolgoprudny 141700, Russia}
\affiliation{Institute for Theoretical and Applied Electrodynamics, Russian Academy of Sciences, Moscow 125412, Russia}

\date{\today}

\begin{abstract}
Partially error-corrected logical executions are expected to become available before fully fault-tolerant quantum computation, but such executions usually take much longer runtime than physical, unencoded ones. 
We formulate zero-noise extrapolation in this regime as a statistical resource-allocation problem in which the physical/logical execution mode is itself an extrapolation design variable. 
In the proposed mixed strategy, one or a few logical circuits provide low-noise anchor points, while cheaper folded physical circuits provide a larger extrapolation lever arm. Within an effective error suppression model $p_L=\gamma p$, we derive Richardson variance prefactors for all-logical and mixed data sets, include folded-circuit runtime accounting, obtain the optimal shot allocation for a prescribed target variance, and state the bias--variance criterion determining when the mixed estimator improves finite-runtime accuracy. We illustrate the mixed-data strategy via simulating dynamics of transverse-field Ising model. 
For error suppression factor $\gamma\lesssim 0.1$ the mixed strategy can significantly(orders-of-magnitude) reduce the runtime needed to reach a fixed estimator variance, as well as provide better mean-square-error estimators in reasonable parameter regions.
\end{abstract}

\maketitle

\section{Introduction}
\label{sec:introduction}

Quantum error correction (QEC) is the long-term route to scalable quantum computing, but its full benefits require a substantial overhead in physical qubits, syndrome extraction, decoding, and control operations \cite{Dennis_2002,Fowler_2012,Devitt_2013,Kivlichan_2020}. Recent experiments show rapid progress toward operating encoded qubits and suppressing logical errors \cite{Zhang_2024,Acharya_2024,He_2025,Wang_2026}. Nevertheless, the intermediate period before full fault tolerance will likely not be described by a clean separation between unprotected physical circuits and perfectly protected logical circuits. A more realistic near-term possibility is that a limited number of partially error-corrected logical executions is available, the residual logical error is still non-negligible, and every logical execution is much more expensive than the corresponding physical execution; this transition regime is also the setting of recent discussions of what can and cannot be ruled out before full fault tolerance \cite{zimboras2025}.

Quantum error mitigation (QEM) offers a complementary route to useful estimates in such a regime. Instead of preserving the full quantum state throughout a computation, QEM protocols use additional noisy circuit executions and classical postprocessing to infer less noisy expectation values \cite{Temme_2017,Li_2017,Kandala_2019,Endo_2018,Endo_2021,Cai_2023}. Zero-noise extrapolation (ZNE) is one of the basic QEM techniques: the same ideal observable is sampled at several effective noise levels and extrapolated to the zero-noise limit \cite{Giurgica_Tiron_2020,Majumdar_2023}. The cost of ZNE is statistical. The extrapolated value is a linear combination of noisy estimates, and the coefficients of this linear combination can strongly amplify finite-shot fluctuations. Optimizing the placement of the noise-scaled points and the number of shots assigned to each point is therefore a central part of the protocol \cite{Krebsbach_2022,Russo_2024}. More generally, QEM is subject to fundamental sampling overheads, so resource accounting must be made explicit \cite{Takagi_2022,Takagi_2023}.

Hybrid use of QEC and QEM is especially natural before full fault tolerance. Previous works have analyzed QEM for encoded gates and error-corrected regimes \cite{Piveteau_2021,Lostaglio_2021}, and have related QEM to error-correcting codes more generally \cite{Suzuki_2022}. Recent strategies include applying ZNE directly to logical qubits by scaling the code distance \cite{Wahl_2023}, applying QEM at the physical layer before decoding \cite{Jeon_2026}, combining clean/logical and noisy/physical qubits in partially error-corrected processors \cite{Bultrini_2023, Koukoulekidis_2023, Dalfavero_2025}, and building mitigation into the QEC machinery through postselection or modified decoding \cite{Smith_2024}. Experiments have also begun to demonstrate QEM on logical qubits and QEM \cite{Zhang_2025}. Still, these developments leave open a complementary resource-allocation question: if logical executions are available but runtime-costly and still noisy, how could we use them to maximize the effect of the hybrid approach?

Here we formulate a complementary view on how to use quantum error correction along with quantum error mitigation and demonstrate that the choice of how many error-corrected data use for quantum error mitigation provides a way for resource optimization.
In particular, in this paper we treat the physical/logical execution mode as a design variable of the ZNE data set. Partial QEC is not used uniformly on every data point; it is allocated selectively to the points where it has the largest statistical value.
The proposed strategy uses logical, partially error-corrected data as low-noise anchors and physical, unencoded data as higher-noise lever-arm points. Within an effective error suppression model in which QEC changes the gate-noise parameter from $p$ to $p_L=\gamma p$, we show that such mixed physical/logical data can significantly reduce the variance amplification of Richardson extrapolation. The resource gain has two origins. First, the larger separation between the low-noise logical anchor and the physical folded points reduces the Richardson coefficients. Second, if logical shots are much slower than physical shots, the all-logical protocol pays the logical runtime cost for every folded circuit, whereas the mixed protocol pays it only for the logical subset. We show how reducing variance via mixed data leads to orders-of-magnitude reduction of runtime required to obtain a zero-noise estimate.
This runtime saving can be particularly valuable for hardware platforms where long physical execution times are themselves a bottleneck \cite{Evered2023,ransford2025,alam2025,granet2026}.

The main results are the following. First, we formulate mixed physical/logical ZNE as a statistical design problem with an effective error suppression model. Second, for Richardson extrapolation we derive the variance prefactors for all-logical and mixed data sets and demonstrate how using mixed data leads to dramatic decrease of runtime under realistic assumptions on parameters. Third, we include the runtime cost of folded circuits and derive the optimal nonuniform shot allocation at fixed target variance and demonstrate runtime reduction of orders of magnitude in mixed-data regime. Fourth, we state the bias--variance condition that determines when the finite-runtime variance reduction also produces a smaller mean-square error (MSE). Finally, we illustrate the mechanism in a digital simulation of transverse-field Ising dynamics.

This paper is organized as follows. In Sec.~\ref{sec:model} we elaborate details on effective error suppression model. In Sec.~\ref{sec:mixed_zne} we consequently unfold the mixed-data method, starting with a geometric illustration of the idea in Sec.~\ref{sec:geometric_picture}, then providing general formulation in Sec.~\label{sec:general_form} and demonstrating it on the Richardson extrapolation example \label{sec:richardson_main}. We then include unitary folding runtime overhead in Sec.~\ref{sec:folded_runtime} and demonstrate in Sec.~\ref{sec:optimal_shots} that mixed data approach is not limited by equal-shots experiments, providing more room to resource management. In Sec.~\ref{sec:bias_variance} we provide a bias-variance trade-off analysis of zero-noise estimators under mixed-data method.

\section{Model, scope, and assumptions}
\label{sec:model}

The mixed estimator is meaningful only after physical and partially error-corrected implementations have been calibrated to a common effective noise coordinate. We assume that both execution modes realize the same target ideal unitary and that the measured expectation value can be parameterized by a single smooth response function $O(\lambda)$, up to corrections smaller than the retained extrapolation terms. Here $\lambda$ denotes an effective total noise strength. For a circuit with $n$ noisy gate locations and a gate-noise parameter $p$, we use the standard coarse coordinate \cite{Cai_2023}
\begin{equation}
\lambda=np.
\label{eq:lambda_np}
\end{equation}

For the logical, partially error-corrected implementation we assume an effective gate-noise parameter
\begin{equation}
p_L=\gamma p,\qquad 0<\gamma\leq 1.
\label{eq:pL_gamma}
\end{equation}
The parameter $\gamma$ is therefore an effective error suppression factor. A motivation of this effective error suppression model is the elementary below-threshold estimate $p_L=c p^2$ for a correction-operation-correction cycle in concatenation code \cite{Devitt_2013}. If $c p_{\rm th}=1$, then $p_L=(p/p_{\rm th})p=\gamma p$ with $\gamma=p/p_{\rm th}$. In the rest of the paper we use Eq.~\eqref{eq:pL_gamma} only as an effective description. A ``logical'' execution in this paper means a partially error-corrected implementation of the same target logical unitary, which still has a non-negligible logical error. All code- and platform-dependent overheads are absorbed into the logical single-shot runtime $\tau_L$ and into $\gamma$. A ``physical'' execution means the corresponding unencoded implementation with single-shot runtime $\tau_p$. We are interested mainly in the regime $\tau_p\ll\tau_L$, although we keep the ratio $\tau_p/\tau_L$ explicit when deriving the optimized-shot results.

We note that our effective error suppression model is by no means a general description of logical error rate under all noise sources of physical quantum computing. Nonetheless, in the regime of partial error correction it is reasonable to assume the use of noise-tailoring methods \cite{Wallman_2016}, which transform coherent errors into incoherent Pauli errors and allow one to control the overall noise model. 
More realistic code-dependent models would change the mapping from device parameters to $\gamma$. The extrapolation formulas below depend only on the calibrated noise coordinates and on the single-shot runtimes assigned to each data point.
We also note that the mixed strategy is a finite-runtime protocol: it can be advantageous when the reduction of estimator variance outweighs the additional extrapolation bias, i.e. when we are limited with the available quantum processor runtime.

\section{Mixed physical/logical extrapolation}
\label{sec:mixed_zne}

\subsection{Geometric two-point motivation}
\label{sec:geometric_picture}

\begin{figure}[t]
    \centering
    \includegraphics[width=0.95\linewidth]{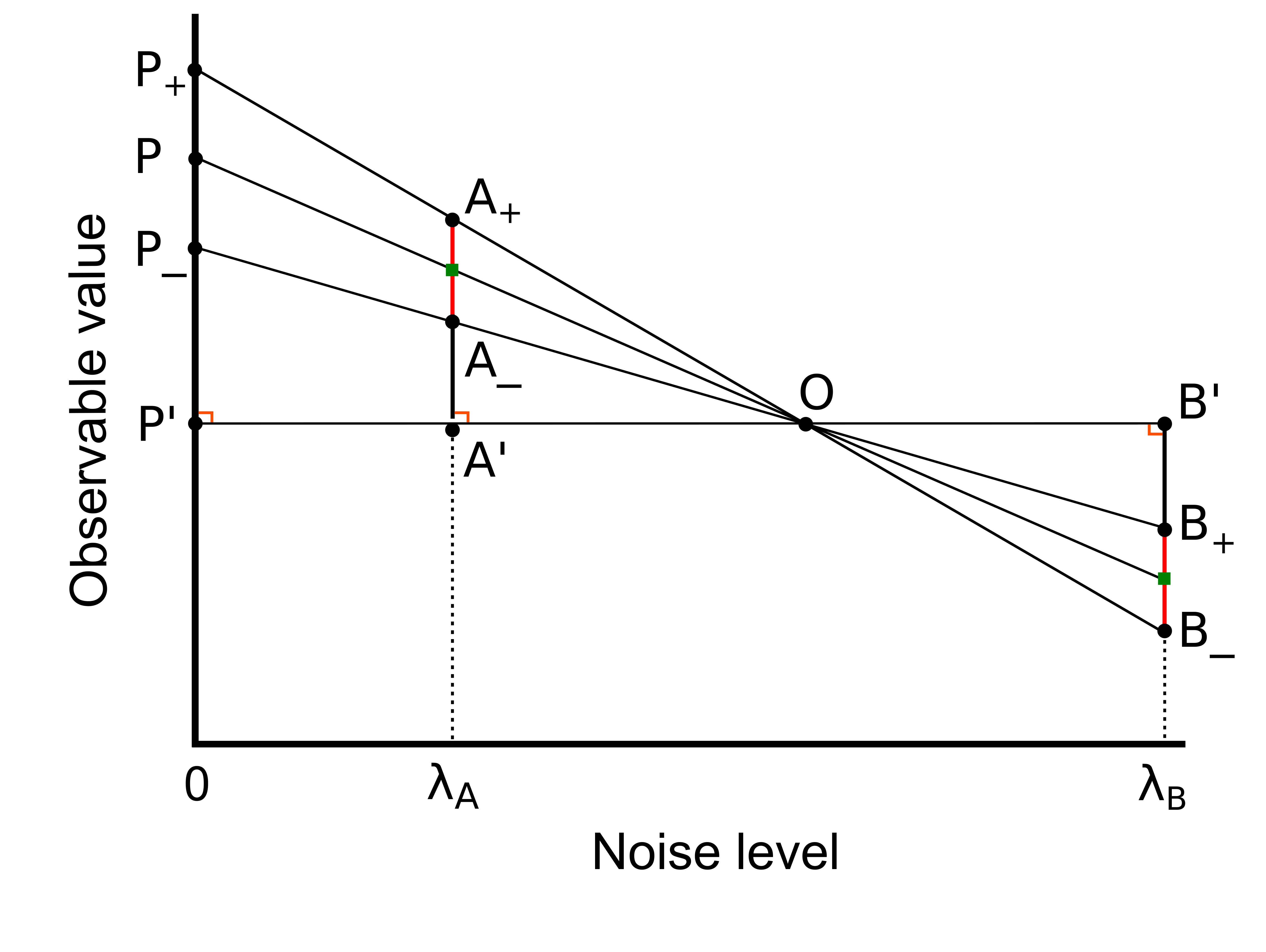}
    \caption{Sketch of linear extrapolation to zero noise for an observable $O(\lambda)$. Green dots represent finite-shot estimates, and red segments represent their statistical uncertainty intervals. The parameters $\lambda_A<\lambda_B$ denote two effective noise levels. The figure illustrates the geometric origin of variance amplification; the quantitative comparisons below use exact linear error propagation.}
    \label{fig:extrapolation_geometry}
\end{figure}

The central idea is shown schematically in Fig.~\ref{fig:extrapolation_geometry}. Instead of obtaining all noise-scaled data points from logical qubits, we use a small number of logical points as low-noise anchors and combine them with physical data at larger noise levels. The logical data reduce the distance from the smallest sampled noise level to zero, while the physical data enlarge the extrapolation baseline at a much smaller runtime cost.

Consider two noise levels $\lambda_A<\lambda_B$ and independent estimates with equal single-point variance $\sigma_s^2/\Np$. The exact two-point estimator is
\begin{equation}
\hat O(0)=\frac{\lambda_B}{\lambda_B-\lambda_A}\hat O(\lambda_A)-\frac{\lambda_A}{\lambda_B-\lambda_A}\hat O(\lambda_B),
\label{eq:linear_estimator}
\end{equation}
with propagated variance
\begin{equation}
\Var[\hat O(0)]=\frac{\sigma_s^2}{\Np}\frac{\lambda_B^2+\lambda_A^2}{(\lambda_B-\lambda_A)^2}.
\label{eq:linear_variance}
\end{equation}
For visual intuition, the interval construction in Fig.~\ref{fig:extrapolation_geometry} gives the conservative width estimate
\begin{equation}
{\rm std}[O(0)]=\frac{\sigma_s}{\sqrt{\Np}}\frac{\lambda_B+\lambda_A}{\lambda_B-\lambda_A},
\label{eq:geometric_std}
\end{equation}
whose derivation is given in Appendix~\ref{app:geometry}. The exact variance formula, Eq.~\eqref{eq:linear_variance}, is used for the Richardson analysis.

For an all-logical two-point design, take $\lambda_A=\gamma np$ and $\lambda_B=2\gamma np$. Equation~\eqref{eq:geometric_std} gives
\begin{equation}
{\rm std}[O(0)]_{\rm L}=3\frac{\sigma_s}{\sqrt{N_{{\rm L}}}},
\label{eq:geo_logical}
\end{equation}
where $N_{\rm L}$ is the number of shots per point. For a mixed design, take the low-noise point from the logical circuit and the high-noise point from the physical circuit: $\lambda_A=\gamma np$ and $\lambda_B=np$. Then
\begin{equation}
{\rm std}[O(0)]_{\rm mix}=\frac{\sigma_s}{\sqrt{N_{{\rm mix}}}}\frac{1+
\gamma}{1-\gamma}.
\label{eq:geo_mixed}
\end{equation}
At equal geometric standard deviation,
\begin{equation}
N_{\rm L}=9\left(\frac{1-\gamma}{1+\gamma}\right)^2N_{\rm mix}\simeq 9N_{\rm mix},\qquad \gamma\ll1.
\label{eq:geo_shots}
\end{equation}
Thus, the mixed two-point design needs fewer shots per point to reach the same extrapolated uncertainty in this motivating estimate. If the physical single-shot runtime is much smaller than the logical one, the same comparison gives
\begin{equation}
\frac{T_{\rm L}}{T_{\rm mix}}\simeq \frac{2N_{\rm L}\tau_L}{N_{\rm mix}\tau_L}=18,\qquad \gamma\ll1,
\label{eq:geo_runtime}
\end{equation}
where $T_{\rm L}$ is the all-logical runtime and $T_{\rm mix}$ is the mixed runtime. This number is only a two-point geometric motivator. The rest of this section gives the exact Richardson variance factors and the corresponding resource accounting.

\subsection{General variance--runtime comparison}
\label{sec:general_form}

Let $K$ denote the polynomial extrapolation order, so that $K+1$ data points are used. For independent estimates with the same single-point variance $\sigma_s^2/\Np$ (where $\sigma_s^2$ is a single-shot variance and $N_{sh}$ is a number of shots), the zero-noise estimator has the generic form
\begin{equation}
\hat O(0)=\sum_{k=0}^{K}\beta_k\hat O(\lambda_k),
\label{eq:generic_estimator}
\end{equation}
and variance
\begin{equation}
\Var[\hat O(0)]=\frac{\sigma_s^2}{\Np}\Fcal,
\qquad
\Fcal=\sum_{k=0}^{K}\beta_k^2.
\label{eq:variance_prefactor}
\end{equation}
The dimensionless quantity $\Fcal$ is the extrapolation-induced variance-amplification factor. It is this factor, together with the single-shot runtimes, that determines the finite-shot resource cost.

Now let us compare two designs of the same polynomial order. In the all-logical design, all $K+1$ points are obtained from partially error-corrected circuits. In the mixed design, we take only one logical data point and the remaining $K$ points are physical. If all data points are assigned equal shot number and equal duration within each execution mode, the equal-variance condition gives
\begin{equation}
\frac{N_{\rm L}}{N_{\rm mix}}=\frac{\Fcal_{\rm L}}{\Fcal_{\rm mix}}.
\label{eq:equal_variance_shots}
\end{equation}
The corresponding runtime ratio is
\begin{equation}
\frac{T_{\rm L}}{T_{\rm mix}}
=\frac{\Fcal_{\rm L}}{\Fcal_{\rm mix}}
\frac{(K+1)\tau_L}{\tau_L+K\tau_p}
\simeq
\frac{\Fcal_{\rm L}}{\Fcal_{\rm mix}}(K+1),
\label{eq:runtime_basic}
\end{equation}
where the last form assumes $\tau_p\ll\tau_L$. Equation~\eqref{eq:runtime_basic} shows the two resource factors. The mixed data set can reduce the extrapolation variance factor, $\Fcal_{\rm mix}<\Fcal_{\rm L}$, and it uses the expensive logical runtime only for the logical anchor.

\subsection{Richardson variance prefactors}
\label{sec:richardson_main}

For Richardson extrapolation, the coefficients in Eq.~\eqref{eq:generic_estimator} are the Lagrange interpolation weights at zero noise,
\begin{equation}
\beta_k=\prod_{l\neq k}\frac{\lambda_l}{\lambda_l-\lambda_k},
\label{eq:richardson_coeff}
\end{equation}
so that
\begin{equation}
\Var[\hat O(0)]=\sum_{k=0}^{K}\beta_k^2\Var[\hat O(\lambda_k)].
\label{eq:richardson_variance}
\end{equation}
Appendix~\ref{app:richardson} gives the derivation.

For all-logical data, we take
\begin{equation}
\lambda_k=M_k\gamma np,
\qquad k=0,\ldots,K,
\label{eq:logical_lambdas}
\end{equation}
where $M_k$ is the digital noise-scaling factor. The variance factor is then
\begin{equation}
\Fcal_{\rm L}
=\sum_{k=0}^{K}\prod_{l\neq k}^{K}\frac{M_l^2}{(M_l-M_k)^2}.
\label{eq:F_logical}
\end{equation}
For the mixed data set, we use a single logical anchor,
\begin{equation}
\lambda_0= M_0\gamma np,
\label{eq:mixed_anchor}
\end{equation}
and physical points
\begin{equation}
\lambda_k=M_k np,
\qquad k=1,\ldots,K.
\label{eq:mixed_physical_lambdas}
\end{equation}
The resulting variance factor is
\begin{widetext}
\begin{equation}
\Fcal_{\rm mix}
=
\prod_{l\neq0}^{K}\frac{M_l^2}{(M_l-\gamma)^2}
+
\sum_{k=1}^{K}\frac{\gamma^2}{(\gamma-M_k)^2}
\prod_{\substack{l\neq k\\ l\neq0}}^{K}\frac{M_l^2}{(M_l-M_k)^2}.
\label{eq:F_mixed}
\end{equation}
\end{widetext}
The first term is the contribution of the logical anchor, and the sum gives the contribution of the physical points. For small $\gamma$, the physical-point coefficients are suppressed, while the anchor coefficient remains close to unity. This is the mathematical origin of the variance reduction.

Tables~\ref{tab:runtime_basic} and \ref{tab:mixed_variance} show numerical values for $M_k=k+1$. The choice $M_k=k+1$ is not essential; it is used only to make the comparison transparent. In practice, the scaling factors depend on the available digital noise-scaling method, such as identity insertion or unitary folding \cite{Giurgica_Tiron_2020}.

\begin{table}[t]
\caption{Runtime ratio $T_{\rm L}/T_{\rm mix}$ from Eq.~\eqref{eq:runtime_basic} for Richardson extrapolation with $M_k=k+1$ and $\tau_p/\tau_L\ll1$. Values larger than one favor the mixed strategy at equal estimator variance.}
\label{tab:runtime_basic}
\begin{ruledtabular}
\begin{tabular}{ccccc}
$\gamma$ & $K=1$ & $K=2$ & $K=3$ & $K=4$ \\
\hline
0.01 & 9.9 & 56 & 270 & 1213 \\
0.1  & 9.0 & 47 & 193 & 569 \\
0.5  & 5.3 & 15 & 27 & 35 \\
0.9  & 2.5 & 4  & 6  & 7
\end{tabular}
\end{ruledtabular}
\end{table}

\begin{table}[t]
\caption{Mixed-data variance factor $\Fcal_{\rm mix}=\Var[\hat O(0)]/(\sigma_s^2/\Np)$ for Richardson extrapolation with $M_k=k+1$. The factor remains close to direct-measurement variance for small $\gamma$ and low-to-moderate polynomial order.}
\label{tab:mixed_variance}
\begin{ruledtabular}
\begin{tabular}{ccccc}
$\gamma$ & $K=1$ & $K=2$ & $K=3$ & $K=4$ \\
\hline
0.01 & 1.01 & 1.02 & 1.02 & 1.03 \\
0.1  & 1.11 & 1.22 & 1.43 & 2.21 \\
0.5  & 1.89 & 3.72 & 10.09 & 36.03 \\
0.9  & 3.98 & 13.51 & 47.85 & 176.85
\end{tabular}
\end{ruledtabular}
\end{table}

For $\gamma\leq0.1$, corresponding to at least an order-of-magnitude effective suppression of the gate-noise parameter, the mixed estimator has a variance factor close to unity for the orders shown. It can therefore reach the same target variance with fewer costly logical executions than the all-logical estimator. The next subsection refines the runtime comparison by accounting for folded-circuit durations.

\subsection{Runtime accounting for folded circuits}
\label{sec:folded_runtime}

Equation~\eqref{eq:runtime_basic} treats all noise-scaled circuits as having the same duration. In unitary folding, however, the circuit at scale $M_k$ is $M_k$ times longer. This matters because in all-logical ZNE even the folded circuits are executed logically, whereas in the mixed design only the anchor is logical.

With folded-circuit durations, the equal-shot runtimes are
\begin{equation}
T_{\rm L}=N_{\rm L}\tau_L\sum_{k=0}^{K}M_k
\label{eq:folded_time_L}
\end{equation}
and
\begin{equation}
T_{\rm mix}=N_{\rm mix}\left(M_0\tau_L+\tau_p\sum_{k=1}^{K}M_k\right).
\label{eq:folded_time_mix}
\end{equation}
Combining Eqs.~\eqref{eq:equal_variance_shots}, \eqref{eq:folded_time_L}, and \eqref{eq:folded_time_mix} gives
\begin{equation}
\frac{T_{\rm L}}{T_{\rm mix}}
=
\frac{\Fcal_{\rm L}}{\Fcal_{\rm mix}}
\frac{\sum_{k=0}^{K}M_k}{M_0+(\tau_p/\tau_L)\sum_{k=1}^{K}M_k}.
\label{eq:runtime_folded_general}
\end{equation}
In the large logical-runtime limit,
\begin{equation}
\frac{T_{\rm L}}{T_{\rm mix}}
\simeq
\frac{\Fcal_{\rm L}}{\Fcal_{\rm mix}}
\frac{\sum_{k=0}^{K}M_k}{M_0}.
\label{eq:runtime_folded_limit}
\end{equation}
This factor can be substantially larger than Eq.~\eqref{eq:runtime_basic} because the all-logical protocol pays the logical runtime cost for several long folded circuits.

\begin{table}[t]
\caption{Runtime ratio $T_{\rm L}/T_{\rm mix}$ from Eq.~\eqref{eq:runtime_folded_limit} for folded-circuit durations and $M_k=k+1$. Values larger than one indicate a runtime advantage of the mixed data set at equal estimator variance under the stated model.}
\label{tab:runtime_folded}
\begin{ruledtabular}
\begin{tabular}{ccccc}
$\gamma$ & $K=1$ & $K=2$ & $K=3$ & $K=4$ \\
\hline
0.01 & 14.8 & 112.1 & 674.1 & 3639.4 \\
0.1  & 13.5 & 93.8  & 482.8 & 1706.7 \\
0.5  & 7.9  & 30.6  & 68.4  & 104.5 \\
0.9  & 3.8  & 8.4   & 14.4  & 21.3
\end{tabular}
\end{ruledtabular}
\end{table}

Table~\ref{tab:runtime_folded} shows the folded-runtime ratio for the same values of $\gamma$ and $K$ as in Table~\ref{tab:runtime_basic}. The advantage grows with extrapolation order because higher-order all-logical ZNE requires more long folded logical circuits.

\subsection{Optimal shot allocation}
\label{sec:optimal_shots}

The previous comparisons assigned the same number of shots to every data point. An optimized experiment would allocate more shots to points with larger extrapolation coefficients and shorter single-shot runtimes. We therefore solve the shot-allocation problem explicitly. This also shows that the mixed-data advantage is not an artifact of equal-shot sampling.

For data point $k$, let $N_k$ be the number of shots and $\tau_k$ the single-shot runtime. At fixed target variance $V$, the optimization problem is
\begin{equation}
\min_{N_k}\sum_{k=0}^{K}N_k\tau_k
\quad
\textrm{subject to}
\quad
\sigma_s^2\sum_{k=0}^{K}\frac{\beta_k^2}{N_k}\leq V .
\label{eq:shot_optimization_problem}
\end{equation}
Using a Lagrange multiplier gives
\begin{equation}
N_k^*=\frac{\sigma_s^2}{V}\frac{|\beta_k|}{\sqrt{\tau_k}}
\sum_{l=0}^{K}|\beta_l|\sqrt{\tau_l},
\label{eq:optimal_shots}
\end{equation}
and the optimized runtime
\begin{equation}
T^*=\frac{\sigma_s^2}{V}
\left(\sum_{k=0}^{K}|\beta_k|\sqrt{\tau_k}\right)^2.
\label{eq:optimal_runtime}
\end{equation}
The same formula can be inverted to give the minimum variance reachable at fixed wall-clock time $T$,
\begin{equation}
V^*(T)=\frac{\sigma_s^2}{T}
\left(\sum_{k=0}^{K}|\beta_k|\sqrt{\tau_k}\right)^2.
\label{eq:optimal_variance_fixed_time}
\end{equation}

For the all-logical design,
\begin{equation}
|\beta^{\rm L}_k|=\prod_{l\neq k}\frac{M_l}{|M_l-M_k|},
\qquad
\tau^{\rm L}_k=M_k\tau_L.
\label{eq:logical_coeff_runtime}
\end{equation}
For the mixed design,
\begin{equation}
|\beta^{\rm mix}_0|=\prod_{l\neq0}\frac{M_l}{|M_l-\gamma|},
\qquad
\tau^{\rm mix}_0=M_0\tau_L,
\label{eq:mixed_anchor_coeff_runtime}
\end{equation}
and for $k=1,\ldots,K$,
\begin{equation}
|\beta^{\rm mix}_k|=
\frac{\gamma}{|M_k-\gamma|}
\prod_{\substack{l\neq k\\l\neq0}}\frac{M_l}{|M_l-M_k|},
\qquad
\tau^{\rm mix}_k=M_k\tau_p.
\label{eq:mixed_physical_coeff_runtime}
\end{equation}
Substitution into Eq.~\eqref{eq:optimal_runtime} yields the optimized runtime ratio
\begin{widetext}
\begin{equation}
\frac{T_{\rm L}^*}{T_{\rm mix}^*}=
\frac{\left(\sum_{k=0}^{K}\prod_{l\neq k}\frac{M_l}{|M_l-M_k|}\sqrt{M_k}\right)^2}
{\left(
\prod_{l\neq0}\frac{M_l}{|M_l-\gamma|}\sqrt{M_0}
+
\sum_{k=1}^{K}\frac{\gamma}{|M_k-\gamma|}
\prod_{\substack{l\neq k\\l\neq0}}\frac{M_l}{|M_l-M_k|}
\sqrt{M_k}\sqrt{\tau_p/\tau_L}
\right)^2}.
\label{eq:optimized_runtime_ratio}
\end{equation}
\end{widetext}
The physical points enter the denominator only through $\sqrt{\tau_p/\tau_L}$. Thus, when physical shots are cheap, the optimized mixed experiment concentrates expensive logical runtime in the anchor and uses physical folded circuits mainly to set the extrapolation lever arm.

\begin{figure}[t]
\centering
\includegraphics[width=0.95\linewidth]{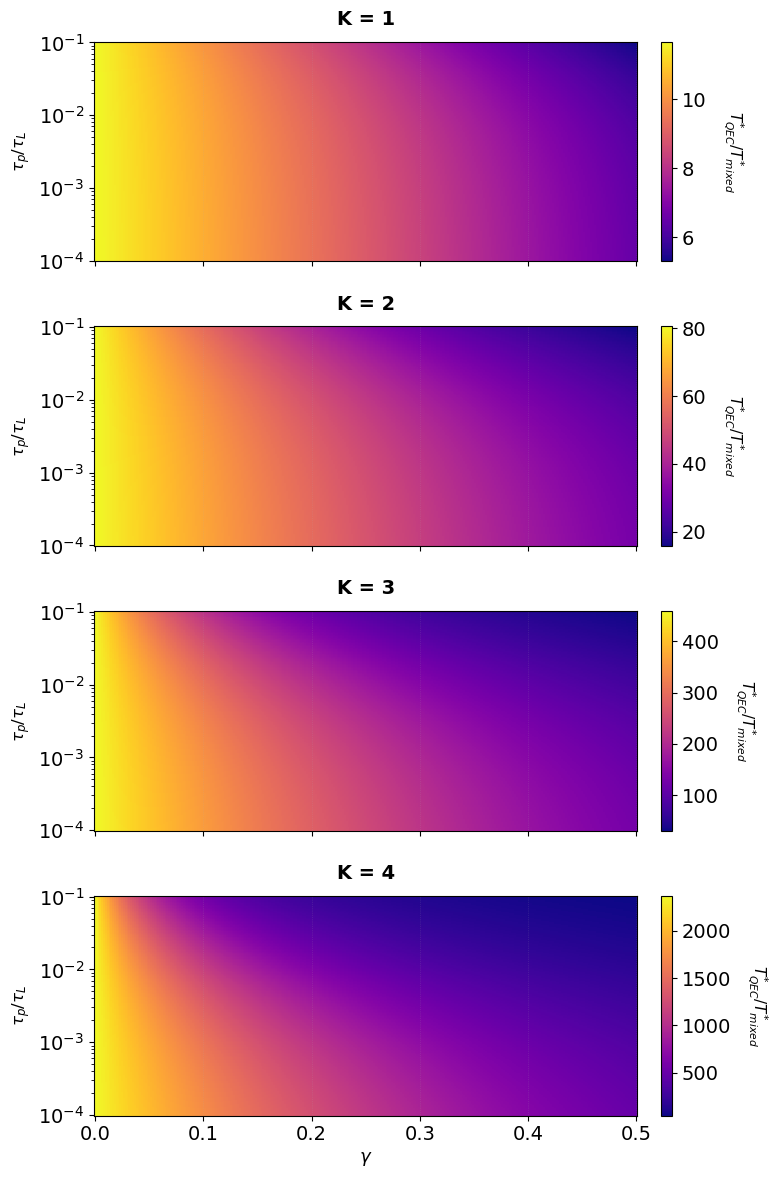}
\caption{Optimized runtime ratio $T^*_{\rm L}/T^*_{\rm mix}$ from Eq.~\eqref{eq:optimized_runtime_ratio} as a function of the error-suppression parameter $\gamma$ and the physical/logical runtime ratio $\tau_p/\tau_L$. The vertical axis is logarithmic. Values larger than one indicate a runtime advantage of the mixed physical/logical strategy at the same target estimator variance after optimal shot allocation.}
\label{fig:runtime_heatmap}
\end{figure}

Figure~\ref{fig:runtime_heatmap} shows this ratio over $\gamma$ and $\tau_p/\tau_L$. The largest advantage appears when $\gamma$ is small and physical shots are much faster than logical ones. In the limit $\tau_p/\tau_L\to0$,
\begin{equation}
\frac{T_{\rm L}^*}{T_{\rm mix}^*}
\to
\frac{\left(\sum_{k=0}^{K}\prod_{l\neq k}\frac{M_l}{|M_l-M_k|}\sqrt{M_k}\right)^2}
{\left(\prod_{l\neq0}\frac{M_l}{|M_l-\gamma|}\sqrt{M_0}\right)^2}.
\label{eq:optimized_runtime_limit}
\end{equation}

\subsection{Bias--variance tradeoff}
\label{sec:bias_variance}

The variance and runtime analysis establishes when the mixed strategy reaches a given statistical precision faster. It does not by itself prove a smaller total error, because the mixed data set includes larger-noise physical points and can therefore have larger extrapolation bias. We now make this tradeoff explicit.

For an analytic response function,
\begin{equation}
O(\lambda)=\sum_{m=0}^{\infty}a_m\lambda^m,
\label{eq:analytic_response}
\end{equation}
Richardson extrapolation of order $K$ removes the first $K$ powers of $\lambda$. The leading interpolation-remainder estimate gives \cite{Atkinson1989}
\begin{equation}
\Bias \simeq a_{K+1}\prod_{k=0}^{K}\lambda_k.
\label{eq:leading_bias}
\end{equation}
For the all-logical and mixed designs above, this gives
\begin{equation}
\Bias_{\rm L}^2=C\gamma^{2(K+1)},
\qquad
\Bias_{\rm mix}^2=C\gamma^2,
\label{eq:bias_scaling}
\end{equation}
where
\begin{equation}
C=a_{K+1}^2(np)^{2(K+1)}\prod_{k=0}^{K}M_k^2.
\label{eq:C_definition}
\end{equation}
The mixed estimator therefore has the larger leading bias for $\gamma<1$ and $K\geq1$. Its advantage, when present, must come from reducing the variance enough to compensate for this bias.

\begin{figure*}[t]
\centering
\includegraphics[width=0.48\linewidth]{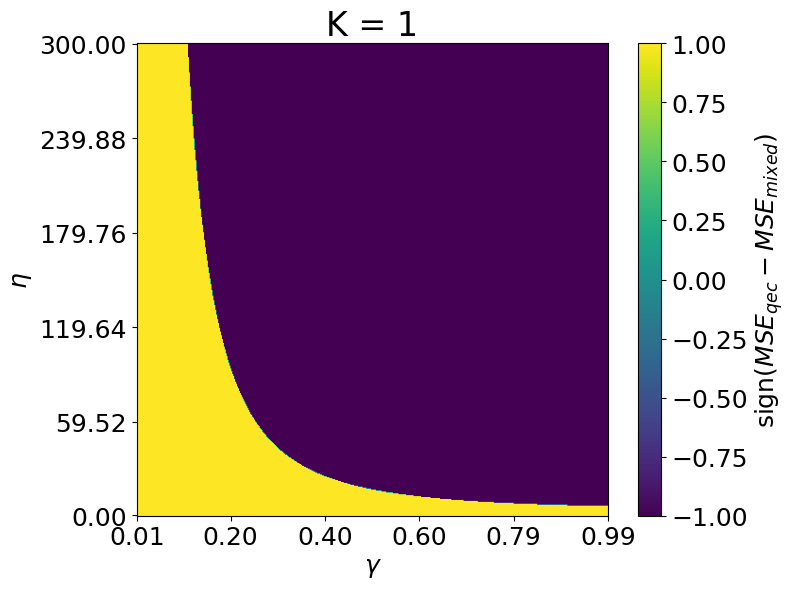}\hfill
\includegraphics[width=0.48\linewidth]{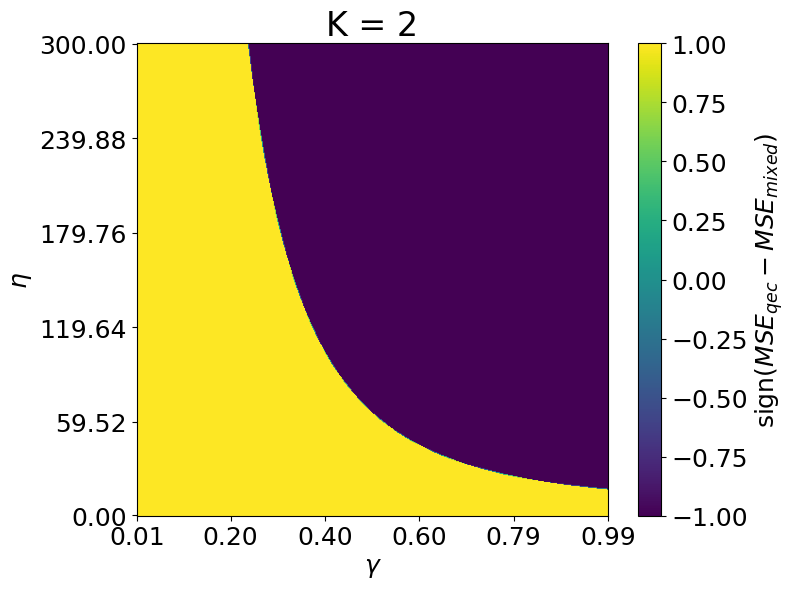}\\[2mm]
\includegraphics[width=0.48\linewidth]{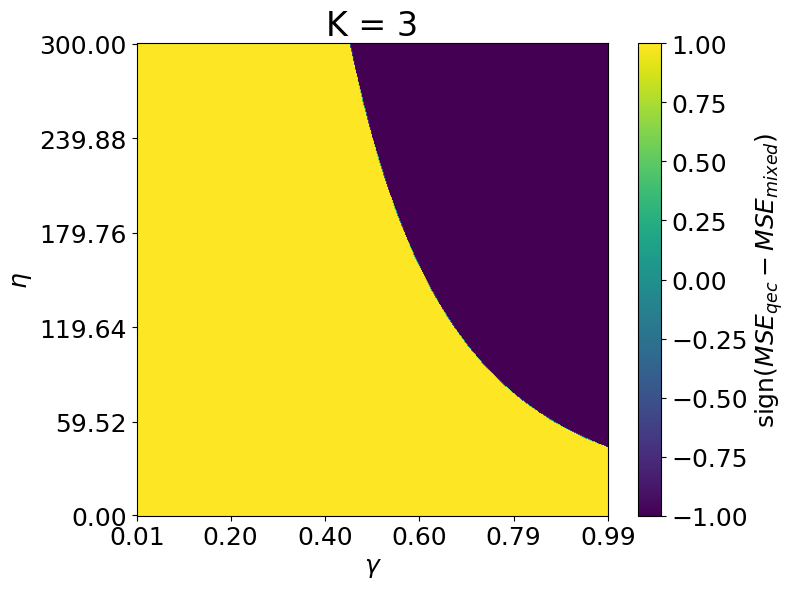}\hfill
\includegraphics[width=0.48\linewidth]{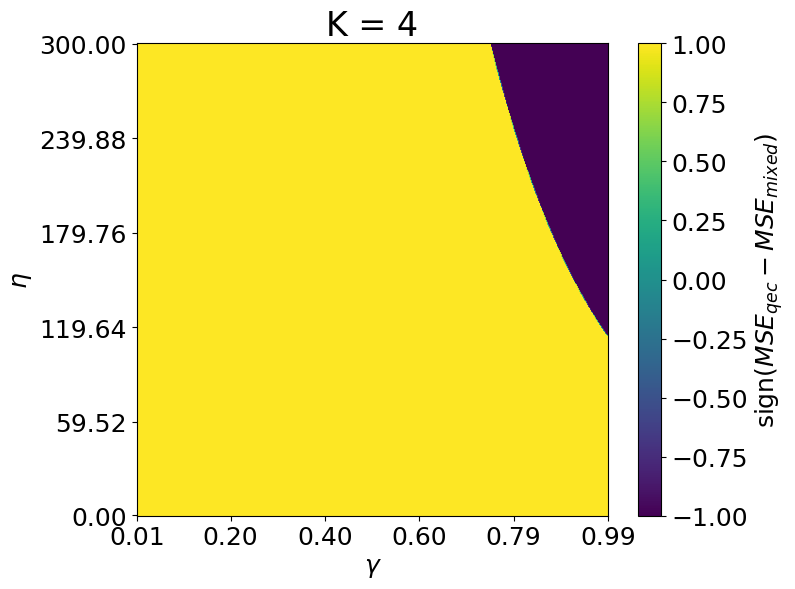}
\caption{Differences of MSE for the two data sets. Each heatmap shows $\operatorname{sign}(\MSE_{\rm L}-\MSE_{\rm mix})$ over $\eta$ and $\gamma$ in the equal-shot normalized comparison. The four panels correspond to Richardson orders $K=1,2,3,4$ from left to right and top to bottom. Yellow regions indicate $\MSE_{\rm L}>\MSE_{\rm mix}$, where the mixed estimator is better; purple regions indicate where the all-logical estimator is better. The figure illustrates a bias--variance tradeoff rather than a universal mixed-data advantage.}
\label{fig:MSE_heatmaps}
\end{figure*}

For equal shot numbers, the MSEs are
\begin{equation}
\MSE_{\rm L}=C\gamma^{2(K+1)}+\frac{\sigma_s^2}{N_{\rm L}}\Fcal_{\rm L},
\label{eq:MSE_logical}
\end{equation}
and
\begin{equation}
\MSE_{\rm mix}=C\gamma^2+\frac{\sigma_s^2}{N_{\rm mix}}\Fcal_{\rm mix}(\gamma).
\label{eq:MSE_mixed}
\end{equation}
If $N_{\rm L}=N_{\rm mix}=\Np$, it is useful to define
\begin{equation}
\eta=\frac{C\Np}{\sigma_s^2}.
\label{eq:eta_definition}
\end{equation}
Then
\begin{equation}
\MSE_{\rm L}=\frac{\sigma_s^2}{\Np}\left[\eta\gamma^{2(K+1)}+\Fcal_{\rm L}\right],
\label{eq:MSE_logical_eta}
\end{equation}
and
\begin{equation}
\MSE_{\rm mix}=\frac{\sigma_s^2}{\Np}\left[\eta\gamma^2+\Fcal_{\rm mix}(\gamma)\right].
\label{eq:MSE_mixed_eta}
\end{equation}
Figure~\ref{fig:MSE_heatmaps} shows the sign of $\MSE_{\rm L}-\MSE_{\rm mix}$ under this normalized equal-shot comparison. The region where mixed data have smaller MSE grows with polynomial order, reflecting the stronger variance amplification of high-order all-logical extrapolation.

At exactly equal variance, the all-logical estimator has smaller leading bias, and hence
\begin{equation}
\MSE_{\rm mix}=C\gamma^2+V>C\gamma^{2(K+1)}+V=\MSE_{\rm L}
\quad (\gamma<1).
\label{eq:equal_variance_mse}
\end{equation}
This is why the mixed strategy should be interpreted as a finite-runtime design. It is useful when the time budget is such that the all-logical variance remains large, while the mixed estimator can obtain many more effective samples.

In general, the mixed strategy is better than all-logical when
\begin{equation}
\MSE_{\rm mix}<\MSE_{\rm L}.
\label{eq:MSE_condition_general}
\end{equation}
Using Eqs.~\eqref{eq:MSE_logical} and \eqref{eq:MSE_mixed}, this gives the criterion
\begin{equation}
\frac{C}{\sigma_s^2}
\leq
\frac{1}{\gamma^2(1-\gamma^{2K})}
\left(
\frac{\Fcal_{\rm L}(K)}{N_{\rm L}}
-
\frac{\Fcal_{\rm mix}(K,\gamma)}{N_{\rm mix}}
\right).
\label{eq:MSE_condition}
\end{equation}
The right-hand side must be positive; otherwise the all-logical design has both lower leading bias and no larger variance contribution.

\begin{figure*}[t]
\centering
\includegraphics[width=0.6\linewidth]{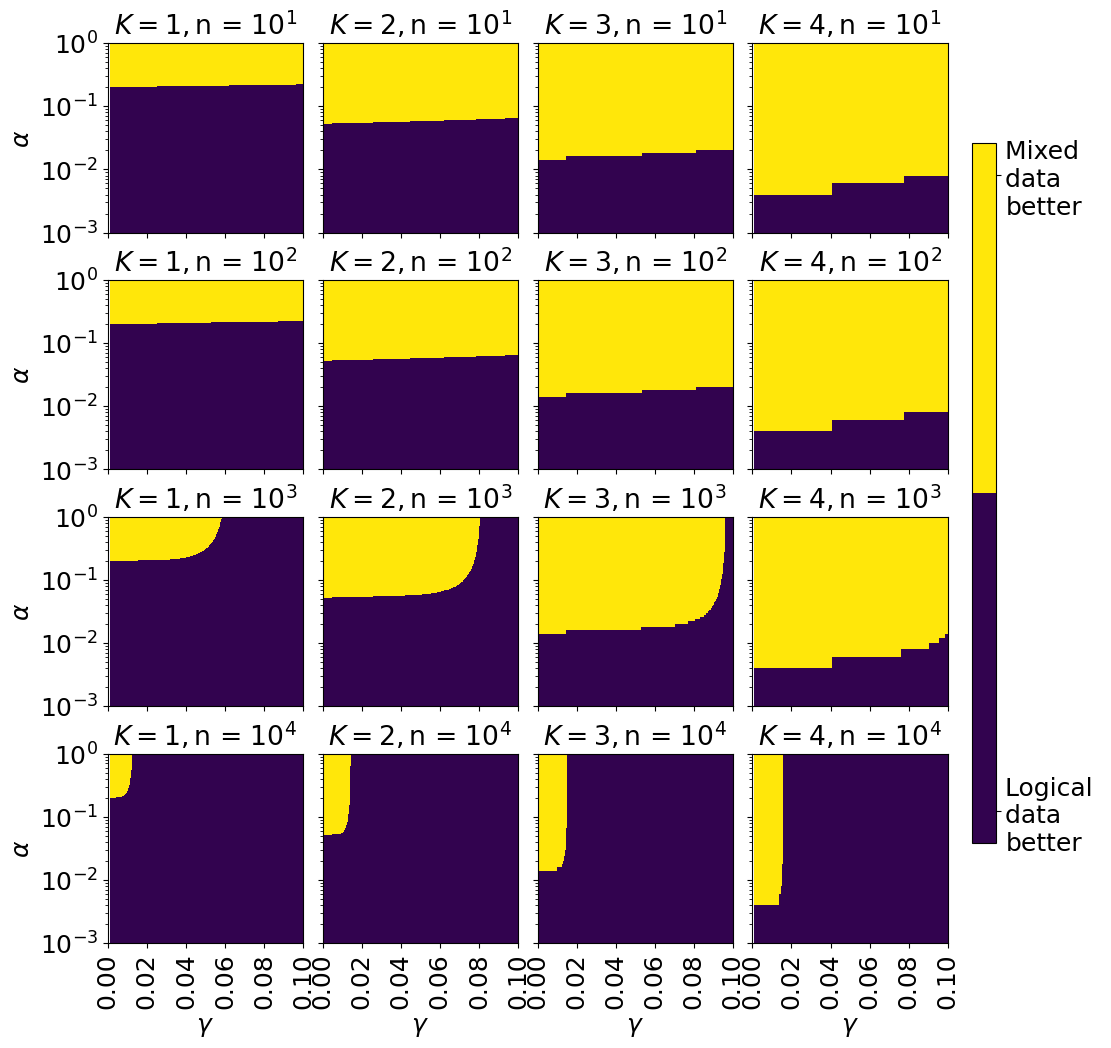}
\caption{Heatmaps for the MSE condition in Eq.~\eqref{eq:MSE_condition}. The vertical axis is logarithmic and represents $\alpha$ in $N_{\rm mix}=\alpha N_{\rm L}$. Yellow regions indicate parameters where the mixed physical/logical estimator has smaller MSE under the leading-bias model; purple regions indicate where the all-logical estimator is better. Rows correspond to different effective gate counts $n$ from $\lambda=np$, and columns correspond to Richardson orders $K$.}
\label{fig:condition_heatmap}
\end{figure*}

To visualize Eq.~\eqref{eq:MSE_condition}, we use the simple response model
\begin{equation}
O(\lambda)=O_0 e^{-\lambda},
\label{eq:exp_response}
\end{equation}
for which
\begin{equation}
a_{K+1}=\frac{(-1)^{K+1}O_0}{(K+1)!}.
\label{eq:a_exp}
\end{equation}
Taking $O_0\leq1$, $p=\gamma p_{\rm th}$, and $p_{\rm th}=0.01$, we evaluate Eq.~\eqref{eq:MSE_condition} for different gate counts $n$ and shot ratios $N_{\rm mix}=\alpha N_{\rm L}$. The result is shown in Fig.~\ref{fig:condition_heatmap}. The yellow regions are finite-runtime regimes where the mixed estimator has smaller leading-order MSE. These regions should not be extrapolated to the singular limit $\alpha\to0$ without keeping the variance term explicit.

The optimized-shot version has the same structure. Using Eq.~\eqref{eq:optimal_variance_fixed_time}, one may write at fixed runtime $T$
\begin{equation}
\MSE_j(T)=B_j^2+\frac{\sigma_s^2}{T}\Rcal_j^2,
\qquad
\Rcal_j=\sum_{k=0}^{K}|\beta_{j,k}|\sqrt{\tau_{j,k}},
\label{eq:fixed_time_MSE}
\end{equation}
where $j\in\{{\rm L},{\rm mix}\}$. The all-logical strategy usually has the smaller leading bias $B_j$, while the mixed strategy can have the smaller runtime--variance factor $\Rcal_j$.

\section{Transverse-field Ising dynamics}
\label{sec:spin_system}

We now demonstrate the mixed physical/logical idea in a proof-of-principle simulation of a six-spin cluster. The analytical results in Sec.~\ref{sec:mixed_zne} are the main contribution of the paper. The numerical example is intended only as a controlled illustration that the variance reduction can translate into improved zero-noise estimates when the bias is not dominant. In this section we use the same number of shots for every data point to keep the comparison transparent; the optimized allocation in Sec.~\ref{sec:optimal_shots} would further reduce the runtime required to reach the same target variance.

\begin{figure}[t]
\centering
\includegraphics[width=0.65\linewidth]{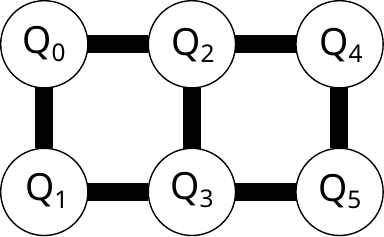}
\caption{Six-spin cluster used in the transverse-field Ising simulation. Edges indicate the coupled spin pairs entering the interaction term of Eq.~\eqref{eq:TFI_hamiltonian}.}
\label{fig:layout}
\end{figure}

The simulated dynamics are governed by the transverse-field Ising Hamiltonian
\begin{equation}
H=h\sum_{i=1}^{N}\sigma_i^X+J\sum_{\langle ij\rangle}^{N}\sigma_i^Z\sigma_j^Z=H_A+H_B,
\label{eq:TFI_hamiltonian}
\end{equation}
with $h=1$ and $J=h/2$. The sum over $\langle ij\rangle$ runs over the coupled pairs in the cluster topology shown in Fig.~\ref{fig:layout}.

We construct the time-evolution operator with a first-order Trotter decomposition using $N_{\rm tr}=80$ steps,
\begin{equation}
U(t)=e^{-it(H_A+H_B)}\approx
\left(e^{-i(t/N_{\rm tr})H_A}e^{-i(t/N_{\rm tr})H_B}\right)^{N_{\rm tr}}.
\label{eq:trotter}
\end{equation}
The one-qubit depolarizing channel is
\begin{equation}
\Phi^{\rm depol}_{1q}(\rho)=(1-p)\rho+\frac{p}{2}I,
\label{eq:depol_1q}
\end{equation}
and the two-qubit depolarizing channel is
\begin{equation}
\Phi^{\rm depol}_{2q}(\rho)=(1-p)\rho+\frac{p}{4}I\otimes I.
\label{eq:depol_2q}
\end{equation}
We use the same $p$ for one- and two-qubit gates, taking it to be of the order of a typical two-qubit gate error. This makes the overall noise level a conservative effective choice.

The target observable is the average magnetization at $T=\pi/2$,
\begin{equation}
\langle M(\pi/2)\rangle=\frac{1}{6}\sum_{j=1}^{6}\langle \sigma^{Z}_j \rangle(\pi/2).
\label{eq:magnetization}
\end{equation}
The effect of partial error correction is modeled by $p_L=\gamma p$, with $\gamma=p/p_{\rm th}$ and $p_{\rm th}=0.01$, an order-of-magnitude threshold value for the surface-code setting used as a reference scale.

We use identity insertion, or global unitary folding, to amplify the noise level digitally \cite{Giurgica_Tiron_2020}. Specifically, we simulate the folded evolutions
\begin{align}
\ket{\psi(T)}_1&=U(T)\ket{\psi(0)},\label{eq:fold1}\\
\ket{\psi(T)}_3&=\left[U(T)U^{\dagger}(T)\right]U(T)\ket{\psi(0)},\label{eq:fold3}\\
\ket{\psi(T)}_5&=\left[U(T)U^{\dagger}(T)\right]^2U(T)\ket{\psi(0)}.
\label{eq:fold5}
\end{align}
These are equivalent in the absence of noise and correspond to noise-scaling factors $M_k=1,3,5$. The effective total noise level is computed as
\begin{equation}
\lambda=\sum_{{\rm gate}\in {\rm circuit}}p_{\rm gate},
\label{eq:lambda_sum}
\end{equation}
where $p_{\rm gate}=p$ for physical circuits and $p_{\rm gate}=p_L$ for logical circuits. The resulting values are listed in Table~\ref{tab:noise_levels}.

\begin{table}[t]
\caption{Effective noise levels used in the six-spin simulation. Labels $1,2,3$ denote logical/error-corrected data, labels $4,5,6$ denote physical/no-error-correction data, and the mixed example uses the set $[1,4,5]$.}
\label{tab:noise_levels}
\begin{ruledtabular}
\begin{tabular}{ccccccc}
Regime & \multicolumn{3}{c}{Logical} & \multicolumn{3}{c}{Physical}\\
Point & 1 & 2 & 3 & 4 & 5 & 6 \\
\hline
$\lambda$ & 0.216 & 0.648 & 1.08 & 2.16 & 6.48 & 10.8
\end{tabular}
\end{ruledtabular}
\end{table}

We collect data for all computational-basis initial states $\ket{\psi(0)}=\ket{i}$, $i=0,\ldots,2^6-1$. Every magnetization value is estimated with $\Np=10^4$ simulated shots. The physical gate error is $p=10^{-3}$, so $\gamma=0.1$. This is the regime where QEC reduces the gate-noise parameter but logical noise remains non-negligible.

\begin{figure}[t]
\centering
\includegraphics[width=0.95\linewidth]{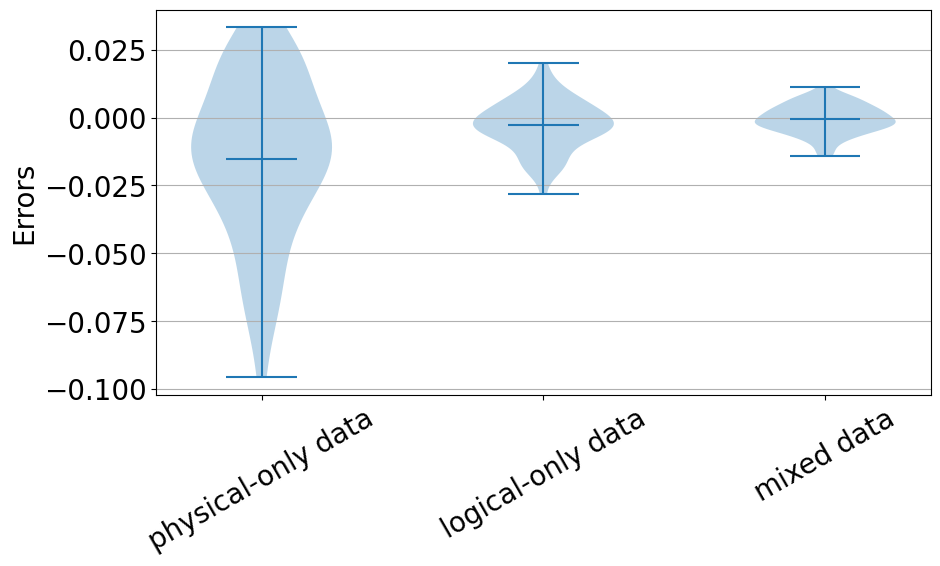}
\caption{Distribution of deviations of zero-noise mean-magnetization estimates from the corresponding noise-free Trotter values. The distribution contains 64 values, one for each computational-basis initial state of the six-spin system. The labels correspond to physical-only, logical-only, and mixed physical/logical data, respectively.}
\label{fig:errors}
\end{figure}

Figure~\ref{fig:errors} shows extrapolation errors over the 64 initial states, where the extrapolation error is the difference between the noise-free Trotter value and the extrapolated zero-noise value. We compare quadratic Richardson extrapolation for three data sets: physical-only $[4,5,6]$, logical-only $[1,2,3]$, and mixed $[1,4,5]$. The physical-only set has predictably larger mean errors and variance. The logical-only set reduces the mean error. In the simulated regime, the mixed set further improves both mean error and variance, consistent with the finite-runtime bias--variance picture above.

\begin{figure}[t]
\centering
\includegraphics[width=0.95\linewidth]{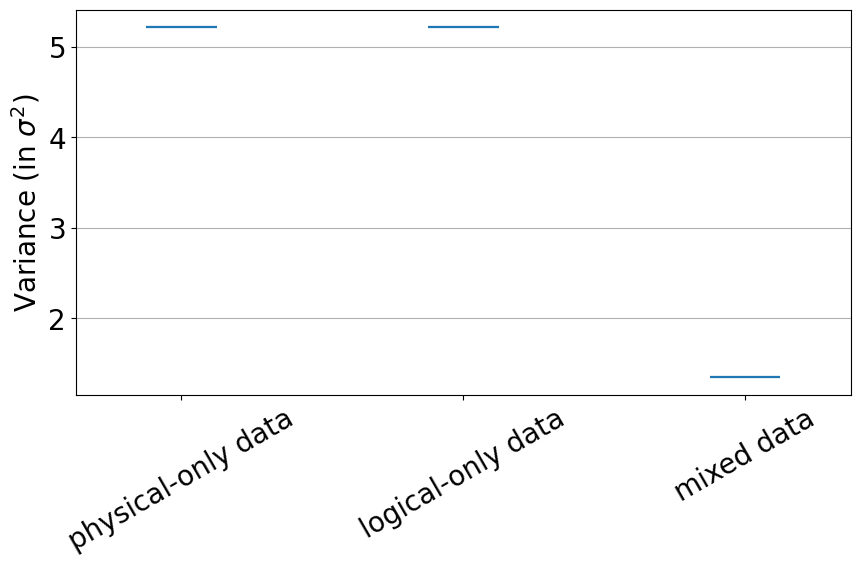}
\caption{Variance-amplification factors of the zero-noise mean-magnetization estimates under equal-shot and equal-single-shot-variance assumptions. The mixed physical/logical data set has a smaller amplification factor because its effective noise levels are more favorable for extrapolation.}
\label{fig:variance_sim}
\end{figure}

Figure~\ref{fig:variance_sim} shows the variance-amplification factors for the same extrapolation estimates. In quadratic Richardson extrapolation with equal measurement uncertainty in each data point, the physical-only set $[4,5,6]$ and the logical-only set $[1,2,3]$ have the same amplification factor because their relative noise spacings are the same. The mixed set has a much smaller factor because the logical anchor and the physical folded points create a wider effective extrapolation baseline.

\section{Discussion and outlook}
\label{sec:discussion}

The analysis above shows that partial QEC can be used as a statistical resource inside ZNE rather than as a uniform replacement for every physical circuit. 
The most favorable regime has three ingredients. First, a meaningful common noise coordinate for physical and logical executions, which can be achieved via noise-tailoring techniques such as a Pauli twirling or dynamical decoupling. Second, logical suppression factor $\gamma$ is below unity, which in essence means a decrease of logical noise from using error-correcting codes. Finally, a large runtime separation $\tau_L/\tau_p$ between physical gates and logical gates, which is true for logical qubits composed of many physical qubits. 
In that regime, one logical low-noise anchor can reduce Richardson variance amplification while the higher-noise physical folded circuits provide the lever arm at much lower runtime cost.

It is useful to distinguish the present protocol from other recent approaches to partial error correction and hybrid QEC–QEM. Refs.\cite{Bultrini_2023, Koukoulekidis_2023} study processors in which a single computation is carried out on a joint register containing both clean/logical and noisy/physical qubits, while Ref.\cite{Dalfavero_2025} uses logical ancillas inside a partially error-corrected processor to reduce the sampling cost of error cancellation via a space–time tradeoff. In contrast, we do not combine logical and physical qubits within one circuit. We run the same target circuit in two separately calibrated execution modes and use the resulting physical and logical data points as a single ZNE data set. The role of partial error correction is therefore not to enlarge the coherent processor directly, but to provide low-noise anchor points that reduce Richardson variance amplification.

The closest ZNE-based comparison is Ref. \cite{Wahl_2023}, which performs ZNE on logical qubits by scaling the error-correction code distance and using the code distance as the noise-scaling knob. Our strategy uses a different knob: the physical/logical execution mode itself. We combine a limited number of partially error-corrected logical executions with cheaper physical folded executions, and we include the physical/logical runtime ratio and folded-circuit durations explicitly in the shot-allocation and MSE analysis. Related approaches mitigate errors inside the QEC stack, for example by applying QEM at the physical layer before decoding \cite{Jeon_2026}, using postselection or modified decoding \cite{Smith_2024}, or directly varying hardware gate errors for noise scaling in error-correction circuits \cite{Zhang_2025}. Our protocol is complementary to these works: it is a statistical design rule for how a finite budget of expensive logical executions should be placed within a ZNE experiment.

Let us stress that the result of this paper is aimed at a transition period of quantum computing towards fault-tolerance, when the computational resources are limited and there is a need for quantum error mitigation. If we can make a quantum state with arbitrarily-suppressed error, we won't need using mixed data and extrapolation at all. Or, if we have no time restriction for running a quantum processor, we may opt for a lowest estimator bias and overcome an increased variance with taking more shots. The mixed-data method is instead aimed at finite-runtime settings in which the number of logical executions is the bottleneck. In such settings, reducing the runtime--variance factor can compensate for the larger bias of the mixed data set.

\section{Conclusion}
\label{sec:conclusion}

We have formulated mixed physical/logical zero-noise extrapolation as a resource-allocation problem for the pre-fault-tolerant regime, which allows for the dramatic decrease of the total runtime under realistic values of main computation parameters. The key point is that we can use logical data (with long runtime) as low-noise anchors while physical executions (with much shorter runtime) can supply the higher-noise points needed for the extrapolation lever arm. Such strategy provides an additional level of resource management, allowing one to use time-heavy logical runs of a quantum processor only for a small part of all runs need to construct a zero-noise estimate.

For Richardson estimators, this construction yields explicit variance-amplification factors. In the effective error suppression model $p_L=\gamma p$, the mixed-data factor remains close to unity for small $\gamma$ and moderate extrapolation order, whereas the all-logical factor can be much larger. Including folded-circuit durations and optimizing shot allocation strengthens the resource interpretation: the all-logical strategy pays the logical runtime cost for every folded circuit, while the mixed strategy concentrates that cost in the logical anchor. As a result, using mixed data leads to order-of-magnitude runtime saving, which is especially valuable for platforms with long physical gate times \cite{Evered2023,ransford2025,alam2025,granet2026} during transition period from NISQ computing to fault-tolerance.

The bias--variance analysis identifies the regime of mixed data method advantage. Mixed data generally lead to a larger extrapolation bias, so the protocol is most useful at finite runtime, when variance reduction outweighs the additional bias. We illustrated mixed data method via simulation of the transverse-field Ising dynamics. Overall, the result provides a simple design principle for hybrid QEC--QEM workflows: partial QEC should not necessarily be spent uniformly across all error-mitigation data, but can be placed where it gives the largest statistical return.

Several extensions are natural. The present work assumes equal single-shot variances for the analytic variance prefactors, although the optimized allocation already allows unequal runtimes. A more detailed hardware model should include unequal measurement variances, reset and measurement times, syndrome-extraction overhead, decoding latency, leakage, and code-dependent logical-noise scaling. It is also interesting to do an experimental demonstration on hardware that can run the same target dynamics in physical and logical modes would provide the most direct test of the protocol.

\section*{Data availability}
All code and data supporting this manuscript are available from the GitHub repository cited in Ref.~\cite{git_ref}.

\section*{Author contributions}
D.B. conceived the study, performed the analysis and simulations, and drafted the manuscript. W.P. supervised the work and contributed to the interpretation and presentation of the results. Both authors reviewed the manuscript.

\appendix

\section{Geometry of extrapolation}
\label{app:geometry}

This appendix derives the conservative geometric uncertainty estimate used as a qualitative motivation in Sec.~\ref{sec:geometric_picture}. The exact propagated variance of the two-point estimator is Eq.~\eqref{eq:linear_variance}; the derivation below concerns the interval-width geometry shown in Fig.~\ref{fig:extrapolation_geometry}.

Let the vertical uncertainty intervals at the two sampled noise levels have lengths $A_+A_-=B_+B_-=2\sigma$, where $\sigma=\sigma_s/\sqrt{\Np}$. The extreme extrapolating lines pass through $(\lambda_A,\tilde O(\lambda_A)-\sigma)$ and $(\lambda_B,\tilde O(\lambda_B)+\sigma)$, or through $(\lambda_A,\tilde O(\lambda_A)+\sigma)$ and $(\lambda_B,\tilde O(\lambda_B)-\sigma)$. Let $P_-P_+$ be the induced interval at $\lambda=0$. From the similarity of the corresponding triangles,
\begin{equation}
\frac{P_-P_+}{A_-A_+}=\frac{P'O}{A'O},
\label{eq:geometry_similarity}
\end{equation}
where $P'O=(\lambda_B+\lambda_A)/2$ and $A'O=(\lambda_B-\lambda_A)/2$. Since $P_-P_+=2\, {\rm std}[O(0)]$ and $A_-A_+=2\sigma$, we obtain
\begin{equation}
{\rm std}[O(0)]=\sigma\frac{\lambda_B+\lambda_A}{\lambda_B-\lambda_A}.
\label{eq:geometry_result}
\end{equation}
Substituting $\sigma=\sigma_s/\sqrt{\Np}$ gives Eq.~\eqref{eq:geometric_std}.

\section{Richardson extrapolation}
\label{app:richardson}

This appendix gives the Richardson formulas used in Sec.~\ref{sec:richardson_main}. For $K+1$ data points $x_k$, $k=0,\ldots,K$, polynomial interpolation gives
\begin{equation}
O(x_k)=\theta_0+\theta_1x_k+\theta_2x_k^2+\cdots+\theta_Kx_k^K.
\label{eq:poly_system}
\end{equation}
The zero-noise value is the interpolation polynomial evaluated at $x=0$,
\begin{equation}
O(0)=\theta_0=\sum_{k=0}^{K}\beta_k O(x_k),
\qquad
\beta_k=\prod_{l\neq k}\frac{x_l}{x_l-x_k}.
\label{eq:appendix_richardson_weights}
\end{equation}
For statistically independent estimates,
\begin{equation}
\Var[\theta_0]=\sum_{k=0}^{K}\beta_k^2\Var[O(x_k)].
\label{eq:appendix_richardson_variance}
\end{equation}
If all input points have the same variance $\sigma^2$, then
\begin{equation}
\Var[\theta_0]=\sigma^2\sum_{k=0}^{K}\left(\prod_{l\neq k}\frac{x_l}{x_l-x_k}\right)^2.
\label{eq:appendix_richardson_equal_variance}
\end{equation}

For all-logical data, $x_k=M_k\gamma np$. Therefore,
\begin{equation}
\frac{x_l}{x_l-x_k}=\frac{M_l}{M_l-M_k},
\label{eq:appendix_logical_ratio}
\end{equation}
and the variance factor is
\begin{equation}
\Fcal_{\rm L}=\sum_{k=0}^{K}\prod_{l\neq k}^{K}\frac{M_l^2}{(M_l-M_k)^2}.
\label{eq:appendix_F_logical}
\end{equation}

For mixed data, $x_0=\gamma np$ and $x_k=M_k np$ for $k=1,\ldots,K$. The anchor contribution is
\begin{equation}
\prod_{l\neq0}\frac{x_l}{x_l-x_0}
=\prod_{l\neq0}\frac{M_l}{M_l-\gamma}.
\label{eq:appendix_anchor_ratio}
\end{equation}
For a physical point $k\neq0$,
\begin{equation}
\prod_{l\neq k}\frac{x_l}{x_l-x_k}
=\frac{\gamma}{\gamma-M_k}
\prod_{\substack{l\neq k\\l\neq0}}\frac{M_l}{M_l-M_k}.
\label{eq:appendix_physical_ratio}
\end{equation}
Squaring and summing the coefficients gives Eq.~\eqref{eq:F_mixed}.

\bibliography{ref}

\end{document}